\documentclass[fleqn,usenatbib]{mnras}

\usepackage{mathptmx}

\usepackage[T1]{fontenc}
\usepackage{ae,aecompl}

\usepackage{graphicx}	
\usepackage{gensymb}
\usepackage{amsmath}	
\usepackage{amssymb}	
\usepackage{numprint}   
\usepackage{footnote}
\usepackage{booktabs}
\usepackage{upgreek}
\usepackage{cleveref}
\usepackage{pdflscape}
\usepackage{float}
\makesavenoteenv{tabular}
\makesavenoteenv{table} 
\crefformat{section}{\S#2#1#3} 
\crefformat{subsection}{\S#2#1#3}
\crefformat{subsubsection}{\S#2#1#3}

\newcommand{\kepler}{\emph{Kepler}}
\newcommand{\ktwo}{\emph{K2}}

\title[``Dipper" Stars in Lupus]{The ASAS-SN Catalog of Variable Stars VIII: \\
\textit{``Dipper'' Stars in the Lupus Star-Forming Region}}

\author[Bredall et al.]{\href{https://orcid.org/0000-0003-3327-6559}{J.~W.~Bredall}$^{1}$,
\href{https://orcid.org/0000-0003-4631-1149}{B.~J.~Shappee}$^{1}$,
\href{https://orcid.org/0000-0002-5258-6846}{E.~Gaidos}$^{2,3}$,
\href{https://orcid.org/0000-0002-6244-477X}{T.~Jayasinghe}$^{4}$,
\href{https://orcid.org/0000-0001-5661-7155}{P.~Vallely}$^{4}$,
\newauthor
K.~Z.~Stanek$^{4,5}$,
C.~S.~Kochanek$^{4,5}$,
\href{http://orcid.org/0000-0002-2592-9612}{J.~Gagn\'e}$^{6,7}$,
K.~Hart$^{1}$,
\href{http://orcid.org/0000-0001-9206-3460}{T.~W.-S.~Holoien}$^{8}$,
\newauthor
\href{https://orcid.org/0000-0003-1072-2712}{J.~L.~Prieto}$^{9,10}$,
\href{https://orcid.org/0000-0002-4284-8638}{J.~Van Saders}$^{1}$
\\
$^{1}$Institute for Astronomy, University of Hawai`i at M\=anoa, 2680 Woodlawn Dr., Honolulu 96822, USA\\
$^{2}$Department of Earth Sciences, University of Hawai`i at M\=anoa, 1680 East-West Rd. Honolulu, HI 96822, USA\\
$^{3}$Kavli Institute for Theoretical Physics, University of California at Santa Barbara, Santa Barbara, CA 93106, USA\\
$^{4}$Department of Astronomy, The Ohio State University, 140 West 18th Avenue, Columbus, OH 43210, USA\\
$^{5}$Center for Cosmology and Astroparticle Physics, The Ohio State University, 191 W.~Woodruff Avenue, Columbus, OH 43210, USA\\
$^{6}$Plan\'etarium Rio Tinto Alcan, Espace pour la Vie, 4801 av. Pierre-de Coubertin, Montr\'eal, Qu\'ebec, Canada\\
$^{7}$Institute for Research on Exoplanets, Universit\'e de Montr\'eal, D\'epartement de Physique, C.P.~6128 Succ. Centre-ville,\\ Montr\'eal, QC H3C~3J7, Canada\\
$^{8}$The Observatories of the Carnegie Institution for Science, 813 Santa Barbara St., Pasadena, CA 91101, USA\\
$^{9}$N\'ucleo de Astronom\'ia de la Facultad de Ingenier\'ia y Ciencias, Universidad Diego Portales, Av. Ej\'ercito 441, Santiago, Chile\\
$^{10}$Millennium Institute of Astrophysics, Santiago, Chile
}
\date{}

\pubyear{2020}

\begin{document}
\label{firstpage}
\pagerange{\pageref{firstpage}--\pageref{lastpage}}
\maketitle
\begin{abstract}
    Some young stellar objects such as T Tauri-like ``dipper" stars vary due to transient partial occultation by circumstellar dust, and observations of this phenomenon inform us of conditions in the planet-forming zones close to these stars. Although many dipper stars have been identified with space missions such as \kepler/\ktwo, ground-based telescopes offer longer term and  multi-wavelength perspectives. We identified 11 dipper stars in the Lupus star forming region in data from the All-Sky Automated Survey for SuperNovae (ASAS-SN), and further characterized these using observations by the Las Cumbres Global Observatory Telescope (LCOGT) and the Transiting Exoplanet Survey Satellite (\emph{TESS}), as well as archival data from other missions. Dipper stars were identified from a catalog of nearby young stars and selected based on the statistical significance, asymmetry, and quasi-periodicity or aperiodicity of variability in their ASAS-SN light curves. All 11 stars lie above or red-ward of the zero-age main sequence and have infrared excesses indicating the presence of full circumstellar disks. We obtain reddening-extinction relations for the variability of 7 stars using our combined ASAS-SN-\emph{TESS} and LCOGT photometry. In all cases the slopes are below the ISM value, suggesting larger grains, and we find a tentative relation between the slope (grain size) and the $\text{K}_\text{s}-[22\:\mu\text{m}]$ infrared color regarded as a proxy for disk evolutionary state.
\end{abstract}

\begin{keywords}
stars: variables: T Tauri, Herbig Ae/Be
\end{keywords}



\section{Introduction}\label{sec:Intro}

Exoplanet science has seen much progress in recent years from both space- and ground-based observations alike (see the reviews by \citealp{Petigura2013}, \citealp{Winn2015}, and \citealp{Perryman2018}). Discoveries by these surveys challenge us to reconcile the layout of the Solar System with what we observe in other planetary systems \citep[e.g.,][]{Mordasini2015, Raymond2018}. One of the biggest obstacles to progress is our limited knowledge of protoplanetary disk structure and the interactions between the disk and the central Young Stellar Object (YSO) \citep[e.g.,][]{Morbidelli2016}. While \textit{Kepler} found many planets within $\lesssim1$ AU of stars, protoplanetary disks are generally not resolved at this scale by current telescopes, including the Atacama Large Millimeter Array ($\sim$5-10 au). Observations of YSO variability offer us a complementary view of phenomena around these stars at separations corresponding to the equivalent Keplerian orbital periods \citep{Cody2018,Hedges2018,Ansdell2019}.

YSO variability is divided into three classes \citep[I--III;][]{Herbst1994}. Each class provides different insight into conditions on or around young stars: Type I, sinusoidal patterns are caused by the rotation of star spots; Type II, episodic ``bursting" events are caused by increases in accretion onto the star; and Type III, dimming events are caused by occulting circumstellar matter. In this paper we focus on Type III variables. UX Orionis stars (UXORs), Herbig Ae/Be stars with occulting circumstellar dust, are a well-known example of this phenomenon \citep[e.g., Co Ori,][]{Davies2018}. Lower mass ($M\lesssim1$ M$_\odot$) T Tauri stars also exhibit dimming due to occulting dust. These were first found in ground-based observations, \citep[e.g., AA Tau,][]{Bouvier1999} but surveys by space telescopes such as \kepler/\ktwo{} have led to the discovery of many more such variables, which have become colloquially known as ``dipper" stars \citep[e.g.,][]{Cody2010, Morales2011, Cody2014, Stauffer2015,Ansdell2016a,Ansdell2018a,Cody2018,Hedges2018}. Unlike UXORs, dipper stars are T Tauri stars which exhibit dimming events ranging from quasi-periodic to completely episodic. The observable difference between UXORs and dippers are the dip timescale, occurrence, and depth, which for dippers are shorter, more frequent, and shallower, respectively, possibly because the inner edge of the dust disk is closer to lower-luminosity dippers \citep{Kennedy2017}. Dippers typically fade from tens of percent to multiple mag for about one day. Dipper variability is believed to be caused by transiting dust, probably associated with the inner regions of a nearly edge-on disk \citep{Stauffer2015, Bodman2017}. Alternative explanations such as exocomets \citep{Boyajian2016} or dusty disk winds \citep{Tamovtseva2008, Bans2012} have also been proposed. Additional studies of dipper stars are needed to distinguish between the differing models.

While space telescopes such as \kepler{} provided high precision and cadence, their temporal baselines are limited to weeks (\textit{Spitzer} and \textit{TESS}) or months (\textit{K2}), and survey only a small fraction of the sky at a time. \emph{CoRoT}, \emph{Spitzer}, and \emph{TESS} conducted individual pointed observations, and the \textit{K2} mission was restricted to the ecliptic plane, and the Taurus and Upper Scorpius star-forming regions. Ground-based observatories such as the Kilodegree Extremely Little Telescope (KELT; \citealp{Pepper2007}) or the Palomar Transient Factory (PTF; \citealp{Law2010}) have been used to survey YSOs, e.g. \citet{Rodriguez2013, Rodriguez2016, Rodriguez2017, Osborn2017, Ansdell2018a} and \citet{vanEyken2011, Findeisen2013}, respectively. KELT had the advantage of all-sky coverage, but was limited to bright stars (8<\textit{V}<12 mag). PTF had a deeper limiting magnitude ($\sim20$ mag), but surveyed only the North American Nebula complex and the 25 Ori region. The successor to PTF, the Zwicky Transient Factory \citep{Graham2019, Bellm2019}, has expanded coverage to the entire night sky visible by Palomar.

Here we perform a search for variable YSOs in the Lupus region using the All-Sky Automated Survey for SuperNovae (ASAS-SN; \citealp{Shappee2014, Kochanek2017}) as part of a larger survey of transient objects. While ASAS-SN has discovered variable YSOs in the past (e.g., \citealp{Holoien2014, Aguilar2017}), the standard pipeline intentionally does not trigger on known objects or low amplitude variability. ASAS-SN offers photometric data of the entire sky from late 2012 to mid 2018 in the \textit{V} band, and from late 2017 to present in the \textit{g} band. The combination of a long baseline and all-sky coverage allows for large surveys of many different types of variability. While the ASAS-SN cadence is slower than \textit{K2} or KELT (typically once per day), ASAS-SN data provides a far wider sky coverage than \textit{K2}, and has a deeper limiting magnitude than KELT. Additionally, ASAS-SN's large longitude coverage allows for observations of the same source by multiple telescopes. As an example, ASAS-SN data has already been successfully used to identify variable stars \citep{Jayasinghe2018, Jayasinghe2019a, Jayasinghe2019b, Pawlak2019, Jayasinghe2019c}, derive period--luminosity relationships for $\delta$ Scuti stars \citep{Jayasinghe2019d}, identify M-dwarf flares \citep{Rodriguez2019}, study contact binaries \citep{Jayasinghe2019e}, and to examine the long-term variability of Boyajian's Star \citep{Simon2018} and compare it with the dipper phenomenon \citep{Schmidt2019}. Finally, ASAS-SN has recently identified a potential new ``Boyajian's Star" analog exhibiting rapid dimming events, nicknamed Zachy's Star \citep{Way2019}.

The constellation Lupus comprises multiple $1$--$3$ Myr-old regions of low-mass star formation near the Scorpius Centaurus OB association (see the review by \citealp{Comeron2008}). At a distance of $189\pm13$ pc, it is one of the closest star-forming regions \citep{Zucker2019}. The Lupus III cloud in particular is home to many well-studied T Tauri stars such as RU Lup and EX Lup. Lupus was not surveyed by \textit{K2}, KELT, or PTF. While there have been several recent studies on disk structure (e.g., \citealp{Ansdell2016b, Ansdell2018b}) and accretion (e.g., \citealp{Nisini2018}) in Lupus YSOs, no dippers have been identified.

Here we use ASAS-SN observations to identify and classify Lupus variable stars in \cref{sec:IdentifyingDippers}. We discuss follow-up observations of dipper sources in \cref{sec:Followup}, and investigate their infrared excesses and dust properties using multi-band photometry in \cref{sec:analysis}.


\section{Identifying Dipper Candidates}\label{sec:IdentifyingDippers}

As the starting point of this investigation, we compiled a catalog of all known candidate members of young associations, open clusters and moving groups within 150 pc based on the member lists of \citet{Torres2008}, \citet{Malo2013}, \citet{Riedel2017}, and \citet{Gagne2018a}, complemented by \citet{Zuckerman2019}, \citet{Furnkranz2019}, \citet{Oh2017}, \citet{Faherty2018}, \citet{Meingast2019}, \citet{Tang2019}, \citet{Gagne2018b}, and \citet{Gagne2018c}. Of these sources, 307 are members of the Lupus Star-Forming Region.

    \subsection{ASAS-SN Observations}\label{sec:asas-sn}

    The ASAS-SN network consists of 20 telescopes mounted on 5 fully-robotic mounts located at the Haleakal\=a Observatory, the Cerro Tololo International Observatory, McDonald Observatory, and the South African Astrophysical Observatory. Observations span from late 2012 to mid-2018 in the \textit{V} band, and from late 2017 in the \textit{g} band, providing $\sim800$ epochs per source on average. Each science image consists of three dithered 90 s exposures taken using 14-cm aperture Nikon telephoto lenses and thinned back-illuminated CCDs with $8.\!\!''0$ pixels.

    Images from ASAS-SN are processed by the fully-automated ASAS-SN pipeline using the ISIS image subtraction package \citep{Alard1998, Alard2000}. The IRAF {\tt apphot} package is used with a 2-pixel ($\approx 16.\!\!''0$) aperture to perform aperture photometry on the subtracted images to generate a differential light curve. The same aperture is used to perform photometry in the reference images, and the resulting fluxes are added back to the differential light curve. The photometry is then calibrated using stars in the AAVSO Photometric All-Sky Survey \citep{Henden2015}. The greatest proper motion measured by \textit{Gaia} for the Lupus YSOs is only $53.68\pm0.05$ mas/yr, so there was no need to take the proper motions into account given our temporal baseline.

    There are known systematic offsets between ASAS-SN telescopes \citep{Jayasinghe2018}. To combine data from different telescopes for a given filter, we use the SciPy {\tt interp1d} package \citep{SciPy} to interpolate the most sampled curve. We then calculate the median difference between this interpolated light curve and the dataset with the highest number of contemporaneous points. We add this difference to the second dataset, combine the light curves, and repeat the process until the data from all telescopes are aligned. This results in each source having a \textit{g}-band and \textit{V}-band light curve.

    To decrease the likelihood of false positives in our variability search, the light curve of each star is processed as follows. First, we remove data points corresponding to images where the FHWM of the point spread function is greater than the 90th percentile value of each telescope. Next, non-detections are omitted from statistical calculations. Lastly, we remove the top and bottom 1\% of detections to clip potential outliers.

    Stars brighter than $10.5$ mag in $V$ and $11$ mag in $g$ are at risk of saturation in ASAS-SN. Approximately 40\% of the stars in our target list are saturated. We removed these brighter sources from our sample. Inconsistent reference fluxes between telescopes yield discrepant magnitude ranges during variable events. This most often took the form of one telescope reporting a large range of magnitudes while simultaneous observations from another appear to ``flatline" by comparison. Any sources that noticeably demonstrated this pattern were omitted. Applying these cuts reduces our sample size from 307 to 177. The \textit{V}-band curves of all remaining sources are processed and analyzed as outlined in \cref{sec:search_var}. Our selection process is summarized in Table \ref{tab:cuts}.

\begin{figure*}
    \centering
    \includegraphics[width=0.8\textwidth]{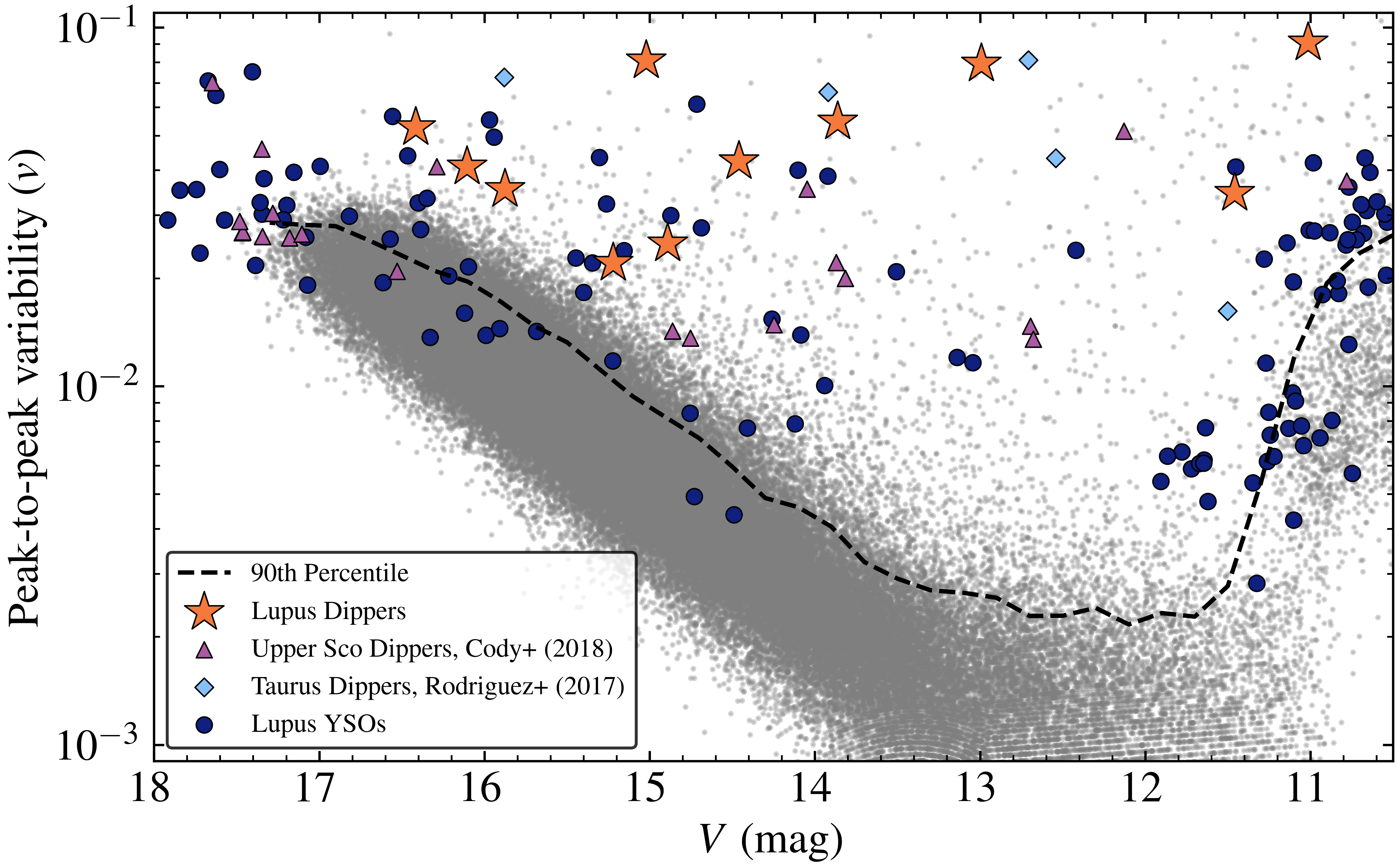}
    \caption{Peak-to-peak variability $v$ as a function of ASAS-SN \textit{V} for our sample of YSOs. Stars flagged as dippers in Upper Scorpius \citep[Table~2 of][]{Cody2018} and Taurus \citep[Table~3 of][]{Rodriguez2017} that were present in our catalog are also shown. The dashed line corresponds to the 90th percentile of $v$ at a given magnitude based on the ASAS-SN \textit{V}-band light curves of about 115,000 randomly selected field stars. Notice how saturation increases the average $v$ at magnitudes brighter than $\sim11.5$. YSOs above the dashed line are considered variable. Out of the $177$ Lupus YSOs analyzed, $91$ pass this criterion for variability. The eleven Lupus dippers described in \cref{sec:classifications} are marked by orange stars.}
    \label{fig:p2p}
\end{figure*}

    \begin{figure*}
        \centering
        \includegraphics[width=0.8\textwidth]{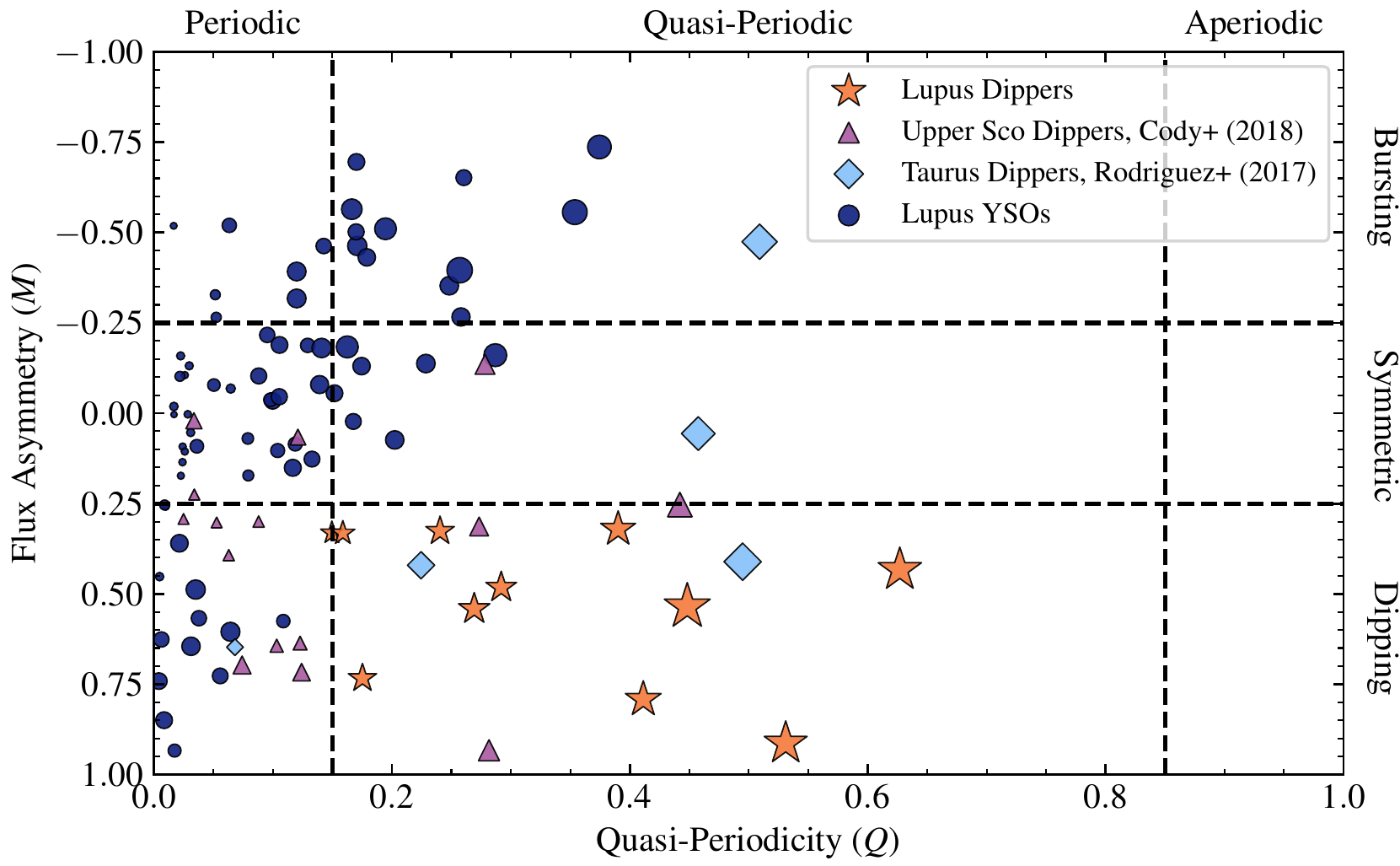}
        \caption{Flux Asymmetry vs Quasi-Periodicity for our sample of variable YSOs. The Flux Asymmetry $M$ is a measure of variability that is predominantly brightening ($M<0$), dimming ($M>0$), or symmetric ($M=0$). The Quasi-Periodicity $Q$ measures whether the variability is more periodic ($Q=0$) or stochastic ($Q>0$). The $Q-M$ space is divided into nine distinct regions to classify variability \citep{Cody2018}. Stars in Lupus that are in the Quasi-Periodic Dipper region are shown with orange stars. The area of each point is proportional to the peak-to-peak variability metric $v$. Stars flagged as dippers in Table~2 of \citet{Cody2018} and Table~3 of \citet{Rodriguez2017} that were present in our sample are also shown.}
        \label{fig:qm}
    \end{figure*}

    \subsection{Searching for Variability}\label{sec:search_var}

    In order to quantify variability, we use the peak-to-peak variability metric
    \begin{equation}
        v=\frac{(m_i-\sigma_i)_\text{max}-(m_i+\sigma_i)_\text{min}}{(m_i-\sigma_i)_\text{max}+(m_i+\sigma_i)_\text{min}},
    \end{equation}
    from \cite{Sokolovsky2017}, where $m_i$ is a measured magnitude and $\sigma_i$ is its associated uncertainty. For a comparison sample, we measured $v$ for the ASAS-SN \textit{V}-band light curves of approximately $115,000$ randomly selected field stars. We calculate the 90th percentile of $v$ as a function of magnitude and selected the $91$ out of $177$ Lupus YSOs above this threshold as variable stars and retained them for further scrutiny. Figure \ref{fig:p2p} illustrates our selection of these sources based on $v$.

    \begin{table}
        \centering
        \caption{Summary of cuts applied to Lupus YSO sample.}
        \label{tab:cuts}
        \begin{tabular*}{\columnwidth}{l@{\extracolsep{\fill}}r}
            \hline
            Criteria &Remaining Sample Size\\
            \hline
            Lupus members in YSO Catalog &307\\
            Removal of Saturated Stars &177\\
            $v>90$th Percentile &91\\
            $Q>0.15$ and $M>0.25$ &11\\
            \hline
        \end{tabular*}
    \end{table}

    \subsection{Classifications}\label{sec:classifications}

    Now that we have a selection of variable YSOs, we next classify their variability using the Quasi-Periodicity and Flux Asymmetry statistics defined in \citet{Cody2014}. The Quasi-Periodicity
    \begin{equation}\label{eq:Q}
        Q=\frac{(\sigma^2_\text{resid}-\sigma_\text{phot}^2)}{(\sigma^2_m-\sigma_\text{phot}^2)},
    \end{equation}
    quantifies the degree to which variability is periodic or stochastic. Here $\sigma_\text{phot}$ is the estimated photometric uncertainty, $\sigma^2_m$ is the variance of the original light curve, and $\sigma^2_\text{resid}$ is the variance of the residual light curve after subtracting the dominant periodic signal. For a perfectly periodic signal, the variance of the residual is equivalent to photometric noise, yielding $Q\approx0$.

    To produce the residual curve, we create a window function (Dirac Comb) out of the original light curve, i.e. values of 1 when an observation takes place and values of 0 elsewhere. We utilize Astropy {\tt LombScargle} \citep{AstropyLombScargle1, AstropyLombScargle2} to find the periodograms of both the original curve and the Dirac Comb window function. We find all frequencies in the window function with a power greater than two standard deviations above the mean and ignore any peaks in the periodogram of the light curve within 0.03 Hz of the peaks in the window function. The curve is then phased using the remaining period with the highest power. We boxcar smooth this folded light curve with a window size of $25\%$ of the period and subtract the smoothed model from the folded curve. We define $\sigma_\text{resid}$ as the standard deviation of this residual curve.

    The second metric to classify variability is Flux Asymmetry
    \begin{equation}
        M=\frac{\left<m_{10\%}\right>-m_\text{med}}{\sigma_m},
    \end{equation}
    where $\left<m_{10\%}\right>$ is the mean of the top and bottom $10\%$ of magnitude measurements, $m_\text{med}$ is the median of all measurements, and $\sigma_m$ is the standard deviation of the light curve. $M$ determines whether the variability is predominantly brightening ($M$ < 0) or predominantly dimming ($M$ > 0).

    \begin{figure*}
        \centering
        \includegraphics[width=0.98\textwidth]{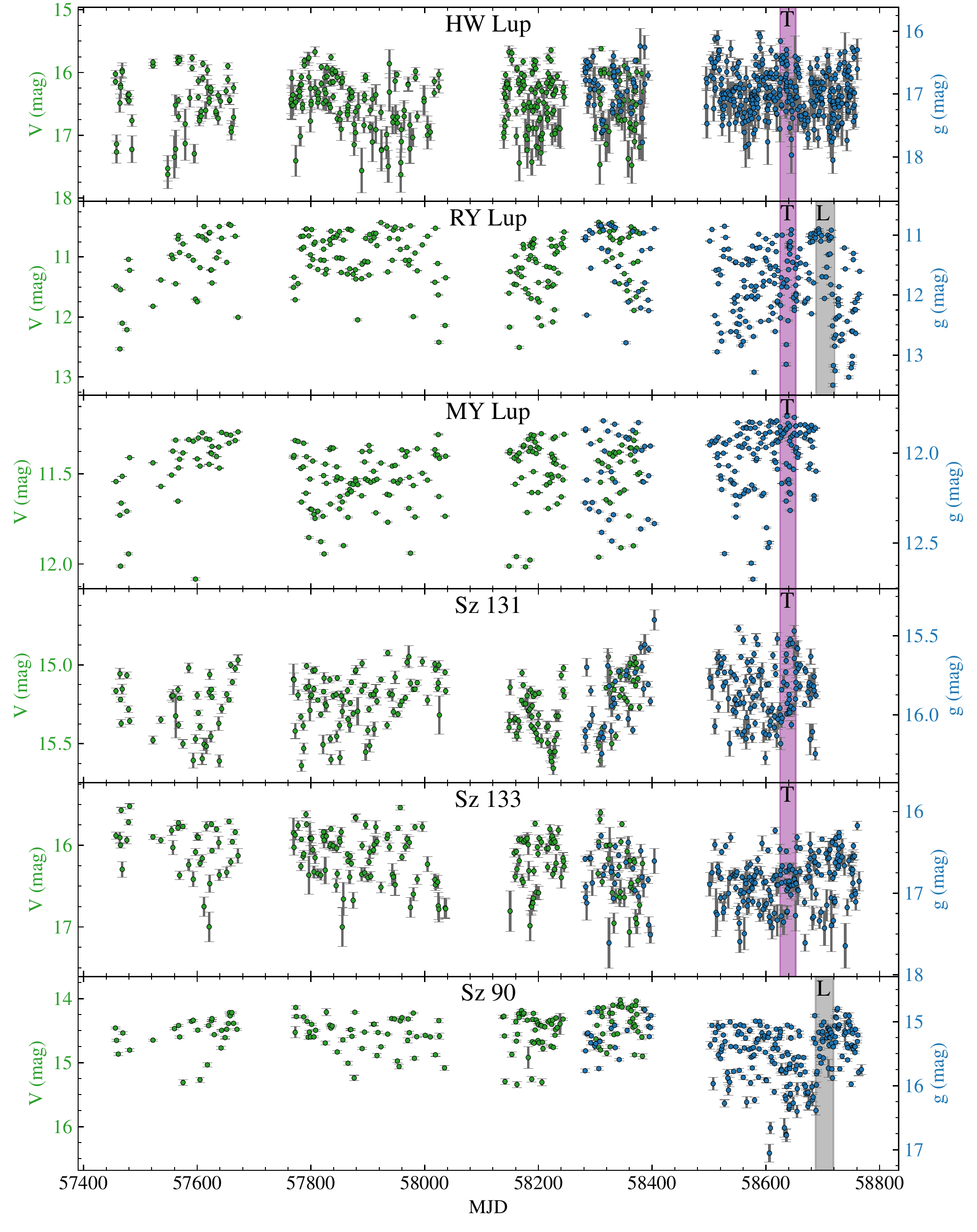}
        \caption{ASAS-SN light curves for the first six dippers in Lupus, in order of ascending RA. Measurements in the $V$-band are in green, while measurements in the $g$-band are in blue. The purple and gray shaded regions show the TESS (T) and LCOGT (L) observing periods, respectively, if applicable to the system. The vertical offset between the filters was found by interpolating the \textit{V}-band data in the overlapping region. The correction is purely for presentation purposes.}
        \label{fig:dippers1}
    \end{figure*}
    \begin{figure*}
        \centering
        \includegraphics[width=0.98\textwidth]{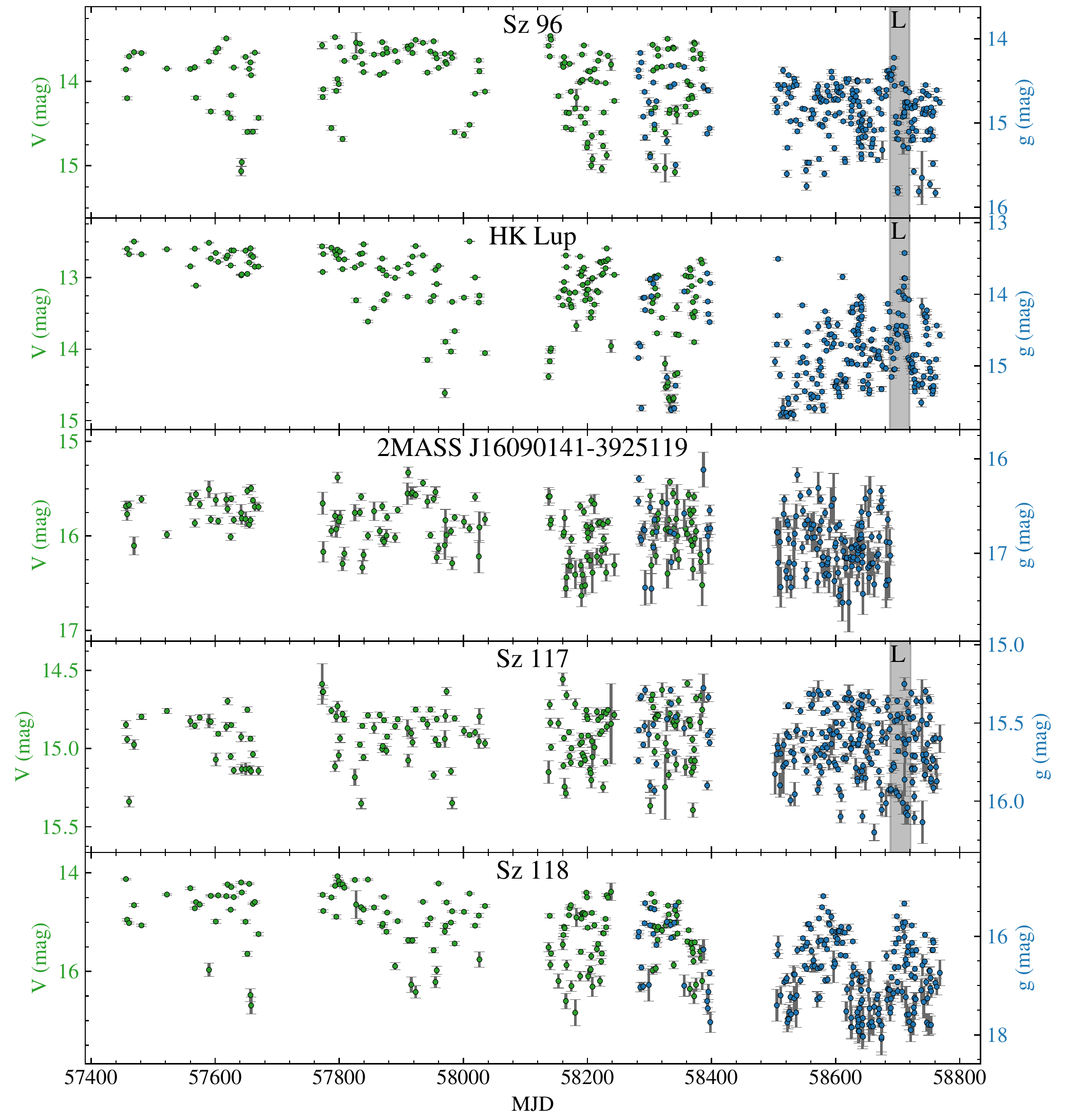}
        \contcaption{ASAS-SN light curves for last five dippers in Lupus, in order of ascending RA. Style is the same as in Figure \ref{fig:dippers1}.}
        \label{fig:dippers2}
    \end{figure*}

    We calculate $Q$ and $M$ for the \textit{V}-band data to separate the YSOs by variability type. The results are shown in Figure \ref{fig:qm}. The $Q-M$ space is divided into nine regions following \citet{Cody2018}. All eleven stars in the Quasi-periodic Dipping region ($0.15<Q<0.85$, $-0.25<M<0.25$) of Figure \ref{fig:qm} were examined in detail, and their light curves are shown in Figure \ref{fig:dippers1}. The designations of all 91 variable Lupus YSOs are included in Table \ref{tab:lup_members}.

     We also performed the same analysis for the ASAS-SN light curves of the dippers identified in \citet{Cody2018} and \citet{Rodriguez2017} that were present in our catalog. These stars are also included in Figures \ref{fig:p2p}, \ref{fig:qm}, and \ref{fig:WISE}. The values of $Q$ and $M$ appear to be inconsistent between ASAS-SN and \textit{K2}, in that there are Upper Scorpius stars identified in \citet{Cody2018} as having dipper-like $Q$ and $M$ values based on \textit{K2} data that have non-dipper values based on ASAS-SN data. We suspect that the discrepant $Q$ and $M$ values are the result of differences in instrumental error and cadence. Ground telescopes such as ASAS-SN sample at timescales comparable to that of individual dipping episodes ($\sim$1 day), whereas \textit{K2} provides a much higher cadence (30 min) and greater precision. A more detailed comparison of $Q$ and $M$ in ASAS-SN as compared to \textit{K2} is left for future work. The ASAS-SN light curves of the flagged dippers (Figure \ref{fig:dippers1}) contain multiple episodic dimming events of up to 2-3 magnitudes, consistent with this variable type.
\begin{landscape}
    \begin{table}
        \caption{Average ASAS-SN $V$ magnitudes, peak-to-peak variability, quasi-periodicity, flux asymmetry, and variability classification for all Lupus YSO members present in our catalog, as well as additional information for quasi-periodic dippers. Names of sources other than quasi-periodic dippers are given as ASAS-SN IDs, generated by their coordinates from the source catalogs (see \cref{sec:IdentifyingDippers}). The classifications are P, QP, and AP for periodic, quasi-periodic, and aperiodic respectively, and D, S, and B for dipper, symmetric, and burster, respectively. A complete list is available electronically.}
        \label{tab:lup_members}
        \begin{tabular}{lcccccrrccc}
        \toprule
            Name & RA & Dec & SpT & V & $v$ & \multicolumn{1}{c}{Q} & \multicolumn{1}{c}{M} & Designation & $A_g/A_T$ & $A_g/A_i$\\
            & deg & deg & & mag & & & & & &\\
        \midrule
            HW Lup                & 236.323  &   $-$34.308 & M1$^\text{a}$    & 16.416 & 0.053 &    0.390 &    0.321 & QPD & $4.791\pm0.018$ &              ---\\
            RY Lup                & 239.868  &   $-$40.364 & G8$^\text{b}$    & 11.015 & 0.091 &    0.448 &    0.536 & QPD & $1.331\pm0.001$ & $1.405\pm0.020$ \\
            MY Lup                & 240.186  &   $-$41.925 & K0$^\text{c}$    & 11.460 & 0.035 &    0.175 &    0.734 & QPD & $1.840\pm0.009$ &              ---\\
            Sz 131                & 240.206  &   $-$41.501 & M2$^\text{a}$    & 15.222 & 0.022 &    0.150 &    0.331 & QPD & $2.998\pm0.010$ &              ---\\
            Sz 133                & 240.873  &   $-$41.667 & K2$^\text{a}$    & 16.106 & 0.041 &    0.292 &    0.482 & QPD & $1.784\pm0.048$ &              ---\\
            Sz 90                 & 241.792  &   $-$39.184 & K7-M0$^\text{a}$ & 14.460 & 0.042 &    0.269 &    0.542 & QPD &             --- &  $1.489\pm0.054$\\
            Sz 96                 & 242.053  &   $-$39.143 &              --- & 13.863 & 0.055 &    0.411 &    0.792 & QPD &             --- &  $1.368\pm0.063$\\
            HK Lup                & 242.094  &   $-$39.079 & M0.4$^\text{d}$  & 12.993 & 0.079 &    0.531 &    0.914 & QPD &             --- &  $1.797\pm0.051$\\
            2MASS J1609           & 242.256  &   $-$39.420 & M4$^\text{e}$    & 15.877 & 0.035 &    0.240 &    0.326 & QPD &             --- &              ---\\
            Sz 117                & 242.435  &   $-$39.225 & M2$^\text{a}$    & 14.891 & 0.025 &    0.159 &    0.332 & QPD &             --- &  $1.084\pm0.035$\\
            Sz 118                & 242.453  &   $-$39.188 & K6$^\text{a}$    & 15.021 & 0.081 &    0.627 &    0.433 & QPD &             --- &              ---\\
            J140220.69$-$414451.1 &  210.586 &  $-$41.7475 &  --- & 10.679 & 0.027 &    $-$0.043 &      0.567 &    PD &     --- &     --- \\
            J140600.39$-$411225.2 &  211.502 &   $-$41.207 &  --- & 11.452 & 0.041 &     0.120 &     $-$0.392 &    PB &     --- &     --- \\
            J142048.90$-$474844.6 &  215.204 &  $-$47.8124 &  --- & 11.622 & 0.005 &     0.017 &      0.004 &    PS &     --- &     --- \\
        \bottomrule
        \end{tabular}
                \begin{flushleft}\footnotesize
            $^\text{a}$ \citet{Hughes1994};\,\,\,\,\,\,\,\,\,$^\text{b}$ \citet{Gahm1989};\,\,\,\,\,\,\,\,\,$^\text{c}$ \citet{Alcala2017};\,\,\,\,\,\,\,\,\,$^\text{d}$ \citet{Herczeg2014};\,\,\,\,\,\,\,\,\,$^\text{e}$ \citet{Romero2012}
        \end{flushleft}
    \end{table}
\end{landscape}


\section{Other Observations}\label{sec:Followup}

Obscuration by dust produces reddening as well as extinction, and the relation between these depends on the grain size distribution and composition. Simultaneous multi-band observations of dipping events allow this to be investigated and compared between dipper stars and with the ISM. For example,  monitoring in both Sloan $g$ and $i$ (or equivalently \emph{TESS} $T$) bands allow the relation between the reddening $\Delta(g-i)$ with dimming $\Delta g$ to be measured, with
\begin{equation}\label{eq:color-mag}
    \frac{\Delta(g-i)}{\Delta g}=1-\frac{A_i}{A_g}
\end{equation}
where $A_g$ and $A_i$ (or $A_T$) are the inferred extinction coefficients for the dust in the Sloan $g$ and $i$ (or \emph{TESS}) bands (see \cref{sec:Dust}). To perform a higher-precision examination of individual dimming events and compare with contemporaneous ASAS-SN measurements, we utilize the observations made by \textit{TESS} during Sector 12 and the Las Cumbres Observatory Global Telescope \citep[LCOGT;][]{Brown2013} 0.4 m telescopes in the $g$ and $i$ bands.

    \subsection{\textit{TESS}}\label{sec:TESS}

    We extracted \textit{TESS} light curves for five of the dipper candidates: HW Lup, RY Lup, MY Lup, Sz 131, and Sz 133. These observations were obtained during Sector 12, near the end of \textit{TESS}'s initial survey of the Southern Hemisphere. The other dipper candidates were not observed by \textit{TESS}.

    We produce light curves from the \textit{TESS} observations using an image subtraction pipeline optimized for use with the full-frame images (FFIs). This pipeline is very similar to that used to process ASAS-SN images and is also based on the ISIS package \citep{Alard1998,Alard2000}. This method has become a standard technique for using \textit{TESS} to study extragalactic transients like supernovae \citep{Vallely2019,Fausnaugh2019} and the exceptional tidal disruption event ASASSN-19bt \citep{Holoien2019}. It has also been previously applied to the study of the EXor variable ESO-H$\alpha$ 99 by \cite{Hodapp2019}.

    For each source, we construct a reference image using 750-pixel-wide postage stamps cut from the first 100 Sector 12 FFIs of good quality. When building the reference images, we exclude FFIs with PSF widths or sky background levels above average for the sector as well as any FFIs associated with non-zero data quality flags. Using these reference images, we produce raw differential light curves for the five sources. However, Both \emph{TESS} and ASAS-SN lightcurves these light curves only provide the change in flux relative to the value in the reference image.   Equation \ref{eq:color-mag} evaluated at two epochs $m$ and $n$ can be written as:
    \begin{equation}\label{eq:ref-flux}
        \frac{A_T}{A_g} = - \log_{10}\left(\frac{\Delta f_{T,m}+f_{T,0}}{\Delta f_{T,n}+f_{T,0}}\right)\left[ \log_{10}\left(\frac{\Delta f_{g,m}+f_{g,0}}{\Delta f_{g,n}+f_{g,0}}\right)\right]^{-1}
    \end{equation}
    This shows that the reddening-extinction slope depends not only on the \emph{TESS} and ASAS-SN differential fluxes $df_{T,m}$ and $f_{g,m}$ but also on the corresponding reference fluxes $f_{T,0}$ and $f_{g,0}$.  However, \textit{TESS}'s PSF is about 16 times larger in area than ASAS-SN's, so crowding affects \textit{TESS} reference fluxes more often.

    Crowding is not a problem for RY Lup and MY Lup, which are sufficiently bright ($T\sim10$ mag) and well-isolated that we were able to obtain good measurements of the reference fluxes using aperture photometry. However, HW Lup, Sz 131, and Sz 133 are fainter and have nearby stars of comparable brightness. For these sources we estimated the reference flux using the {\tt ticgen} software package \citep{Ticgen,Stassun2018}. This method yields \textit{TESS} magnitude estimates for HW Lup, Sz 131, and Sz 133 of 12.26, 11.56, and 13.86 mag, respectively. These magnitude estimates were converted into fluxes using a standard instrumental zero point of 20.44 electrons per second \citep{TESSHandbook}. We then added flux to the raw differential light curves such that the median of the HW Lup, Sz 131, and Sz 133 observations matched this estimated reference value. This method has been used previously by \citet{Jayasinghe2019d} when studying $\delta$ Scuti stars with \textit{TESS}. However, we caution that for non-periodic, inherently variable sources such as dippers, estimating the reference flux from {\tt ticgen} should be made with higher-resolution contemporaneous observations with the \textit{TESS} reference image. Because no such observations were made, the values of $1-A_T/A_g$ for HW Lup, Sz 131, and Sz 133 are influenced by the difference between the {\tt ticgen} reported flux and the actual flux of the reference image. Future investigations can resolve this problem with a single epoch of higher resolution, multi-band, ground-based observations during a \textit{TESS} campaign to calibrate the differential flux light curve.

\begin{figure*}
        \centering
        \includegraphics[width=0.96\textwidth]{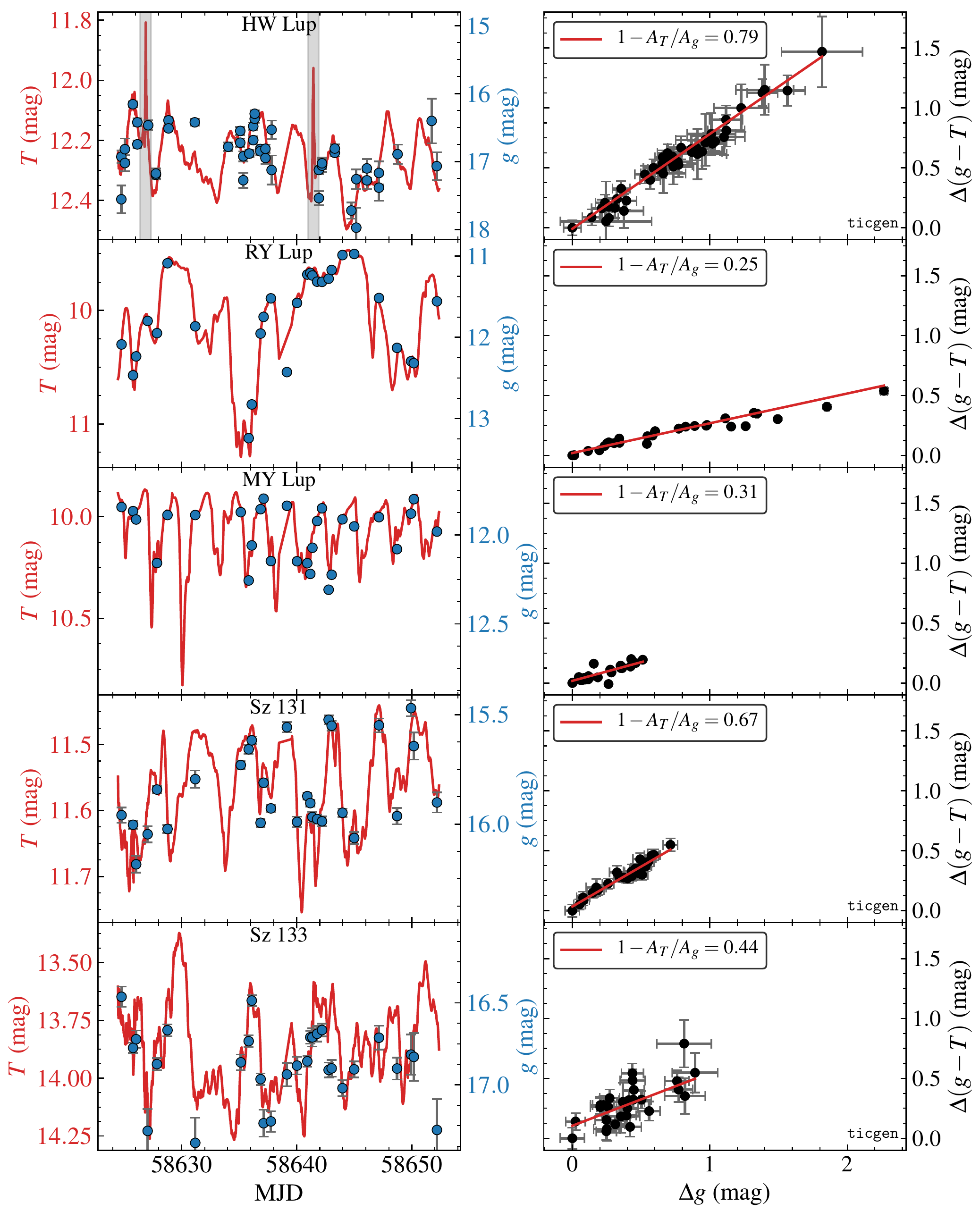}
        \caption{{\it Left Panels:} Simultaneous \textit{TESS} (red) and ASAS-SN $g$ (blue) observations of five of the dipper candidates. The data from ASAS-SN has been vertically scaled to demonstrate the agreement with \textit{TESS} on variability structure, with the right axis reporting the actual measured ASAS-SN magnitudes. The vertical gray regions indicate epochs when scattered Earthshine artifacts are dominant. \textit{TESS} data from these epochs is excluded from the subsequent analysis. {\it Right Panels:} Reddening vs Dimming between ASAS-SN and \textit{TESS}. The slope of the fitted line is equivalent to $1-A_T/A_g$. The value for interstellar dust is $1-A_T/A_g=0.38$. Sources marked with {\tt ticgen} were located in crowded fields and could not have their reference flux measured directly. Consequently the slopes of these sources should be met with skepticism.}
        \label{fig:TESS}
    \end{figure*}

    The resultant \textit{TESS} photometry is shown in Figure \ref{fig:TESS}, along with the simultaneous ASAS-SN \textit{g}-band observations. We note that the two short-duration features visible in the \emph{TESS} light curve of HW Lup around MJD 58627 and MJD 58641 (marked with grey vertical bands) are almost certainly artifacts produced by scattered Earthshine. Excess background light due to scattered Earth and Moon light is a well-known nuisance characteristic of \textit{TESS} observations, and the two spikes in the HW Lup light curve coincide with epochs of documented high background flux (see the Sector 12 \textit{TESS} Data Release Notes\footnote{\url{https://archive.stsci.edu/missions/tess/doc/tess_drn/tess_sector_12_drn17_v02.pdf}} for more details). Aside from these artifacts, the ASAS-SN and \textit{TESS} data sets show generally excellent agreement.

    \subsection{LCOGT}\label{sec:LCOGT}

    \begin{figure*}
        \centering
        \includegraphics[width=\textwidth]{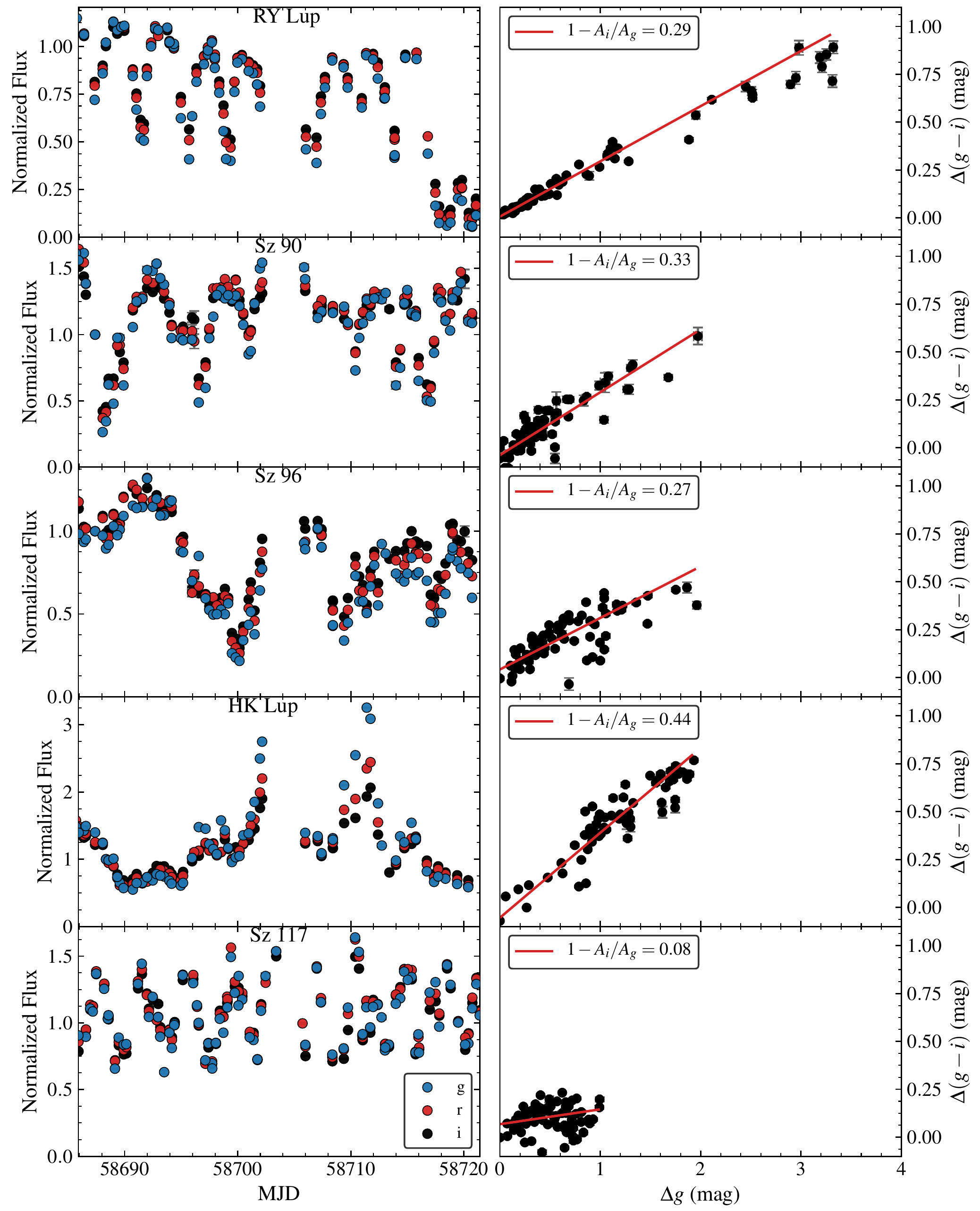}
        \caption{LCOGT SDSS $gri$ photometry of 5 of the dipper stars identified using ASAS-SN. {\it Left Panels:} Normalized light curves. {\it Right Panels:} Reddening vs. Dimming trends. Red lines are least-squares linear fits.}
        \label{fig:lco}
    \end{figure*}

    We also obtained $g$, $r$, and $i$ band observations of five Lupus dippers with the LCOGT 0.4-m telescope network, specifically the nodes at Suderland South Africa, CTIO, and Siding Spring, from 21 July, 2019 to 28 August, 2019. Images were acquired with the SBIG detector. Bias removal, image trimming, flat fielding, and photometry were performed with the {\tt BANZAI} pipeline \citep{McCully2018}. Further analysis of the photometry was performed with custom differential scripts, choosing a set of reference stars with similar magnitudes in each case. RY Lup and Sz 117 were observed for 77 and 80 epochs, respectively, and Sz 90, Sz 96, and HK Lup were observed for 85 epochs (Figure \ref{fig:lco}).



\section{ANALYSIS}\label{sec:analysis}

    \subsection{Dippers in the Color-Magnitude Diagram}\label{sec:hrdiagram}

\begin{figure}
        \centering
        \includegraphics[width=\columnwidth]{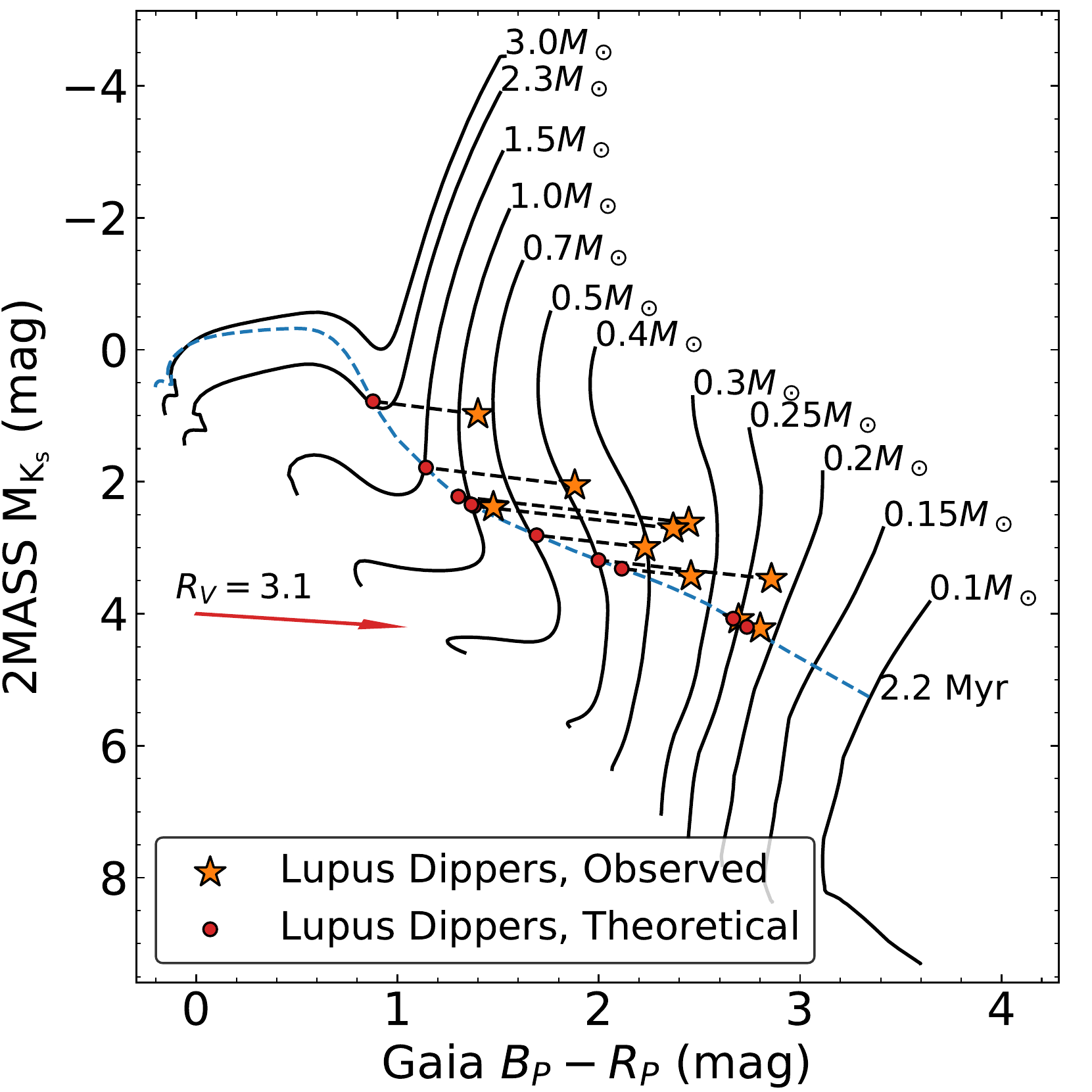}
        \caption{\textit{Gaia} $B_P-R_P$ color vs. absolute $\text{M}_{\text{K}_\text{s}}$ color-mag diagram for ten of the dippers. Absolute magnitudes are obtained using distances from \citet{Bailer-Jones2018}. The isochrone models are from MIST. By assuming an age of 2.2 Myr for the dippers and iteratively applying a reddening correction, we obtain rough estimates for stellar masses.}
        \label{fig:HR}
        \vspace{-1em}
    \end{figure}

    At the age of the Lupus molecular cloud, low-mass T Tauri stars such as these dippers have not reached the zero-age main sequence and should appear above the Zero-Age Main Sequence (ZAMS) in a color-magnitude diagram. We computed absolute magnitudes in the 2MASS $K_s$ band \citep{Skrutskie2006} using parallaxes from \emph{Gaia} DR2 \citep{Gaia2018} and show these with \emph{Gaia} $B_P-R_p$ colors as orange stars in Figure \ref{fig:HR}. Note that Sz 133 lacks a DR2 parallax and does not appear. We compare these with predictions from Isochrones and Stellar Tracks \citep[MIST;][]{MIST1, MIST2, MIST3, MIST4, MIST5, MIST6}. Extinction by interstellar dust is comparatively low in the $K_s$ band relative to \emph{Gaia} $B_P$ or $R_p$ and interstellar reddening will cause a star to move nearly horizontally to the right in this diagram, and so appear further above the ZAMS, giving the impression that they are both younger and less massive. Without spectroscopic follow-up, we cannot robustly estimate the $T_{\rm eff}$ and hence intrinsic $B_p-R_p$ color of these stars to reliably place them in this diagram to constrain their ages and masses. Moreover, the time-averaged visible-wavelength $B_P - R_P$ color reported by DR2 will be affected both by the mean level of \emph{circumstellar} dust as well as spots on these young, active stars, which are not accounted for by the MIST models \citep{Somers2015, Somers2020}. However, we can obtain an approximate estimate of the stellar masses if we \emph{assume a minimum co-eval age}. By adopting 2.2 Myr as the minimum co-eval age of the dipper stars, as well as a reddening relation for the ISM  ($A_{\text{K}_\text{s}}=0.194\,\,\text{E}(B_p-R_p)$; \citealp{Yuan2013}) we can project (dashed black lines and red dots in Figure \ref{fig:HR}) these stars onto the isochrone (dashed blue line) and estimate the minimum reddening and minimum masses of the stars. Most of the dippers have $M\lesssim1\,\text{M}_\odot$ with the notable exception of RY Lup, where we obtain $M\sim2.3\,\text{M}_\odot$ using this method. This is consistent with the picture of dipper stars as low mass T Tauri stars, to be distinguished from their higher mass UXOR Ae/Be-type cousins.

    \subsection{Infrared Excess and Disks}\label{sec:Disks}

\begin{figure*}
        \centering
        \includegraphics[width=0.8\textwidth]{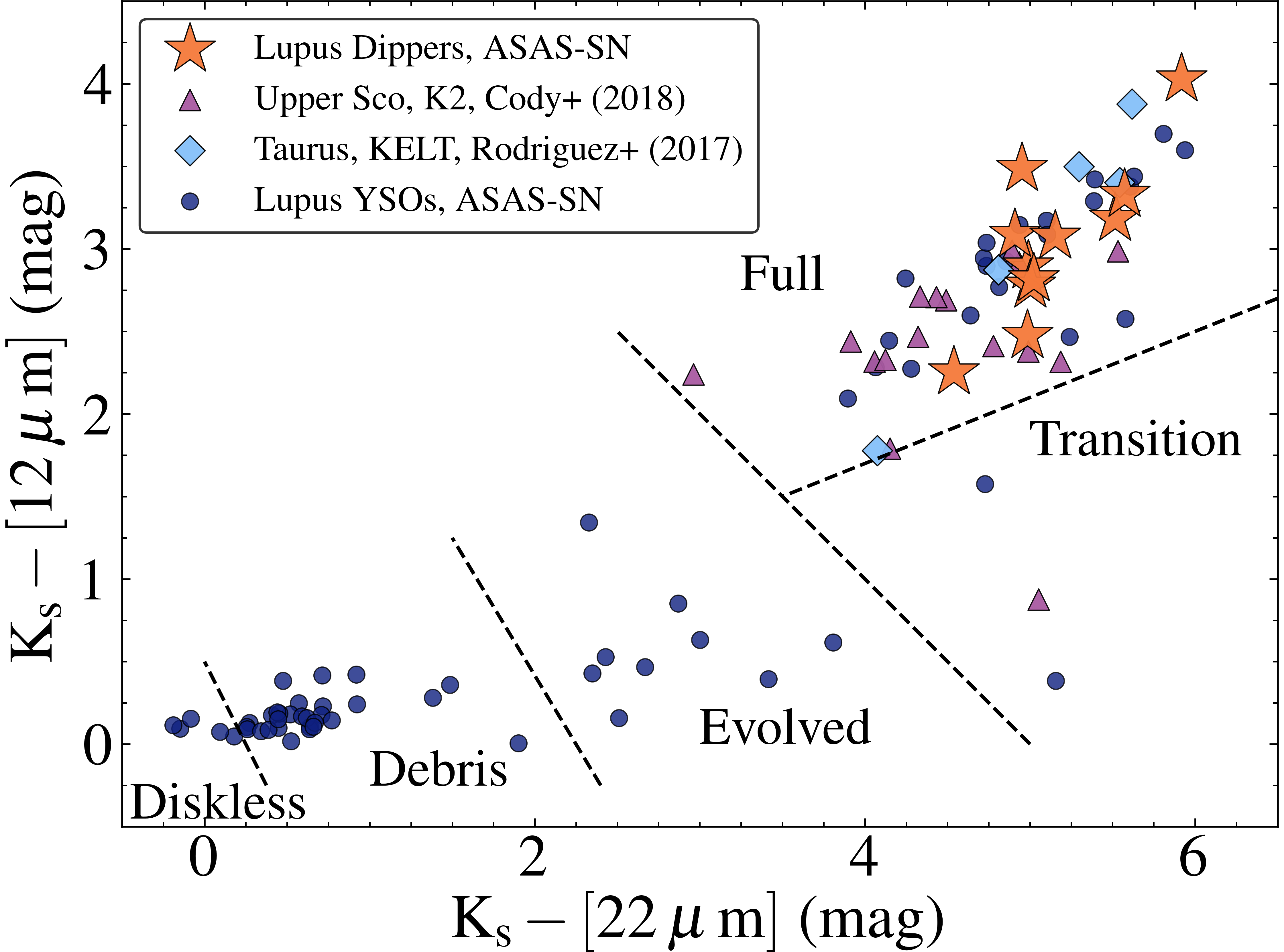}
        \caption{2MASS--\textit{WISE} color--color diagram. Zones characteristic of the different evolutionary stages of disks as defined by \citet{Luhman2012} are demarcated by the dashed lines. The Lupus dippers are labelled alongside other known dippers. Other Lupus stars in our YSO catalog are shown in blue. The infrared excess of the dippers is consistent with dipping being related to the presence of a disk.}
        \label{fig:WISE}
    \end{figure*}

    For young stars, excess emission in the infrared is indicative of the presence of a disk. By comparing a star's absolute flux in the 2MASS $\text{K}_\text{s}$ band, which will be dominated by the photosphere, with their 12 ($W3$) and 22 ($W4$) $\mu$m fluxes from the Wide-field Infrared Survey Explorer \citep[WISE;][]{Wright2010, Mainzer2011}, we can ascertain the presence of a disk and investigate its evolutionary state \citep{Luhman2012}. Figure \ref{fig:WISE} shows the $\text{K}_\text{s}-[12\:\mu\text{m}]$ vs. $\text{K}_\text{s}-[22\:\mu\text{m}]$ colors of the Lupus YSOs along with other reported dippers in Upper Scorpius and Taurus. The boundaries for disk types defined by \citet{Luhman2012} are included. Young stars with ``full" or classical T Tauri disks occupy the upper right hand portion of the figure. As the disk evolves, the star will lose its infrared excess. The relative rates at which $\text{K}_\text{s}-[22\:\mu\text{m}]$ and $\text{K}_\text{s}-[12\:\mu\text{m}]$ decrease are indicative of how and where the disk is evolving. All eleven dipper candidates show infrared excesses consistent with that of full disk-bearing YSOs, along with all but one of the previously reported dippers in our sample. This further supports a connection between the presence of a disk and the dipper phenomenon.

    \begin{figure*}
        \centering
        \includegraphics[width=\textwidth]{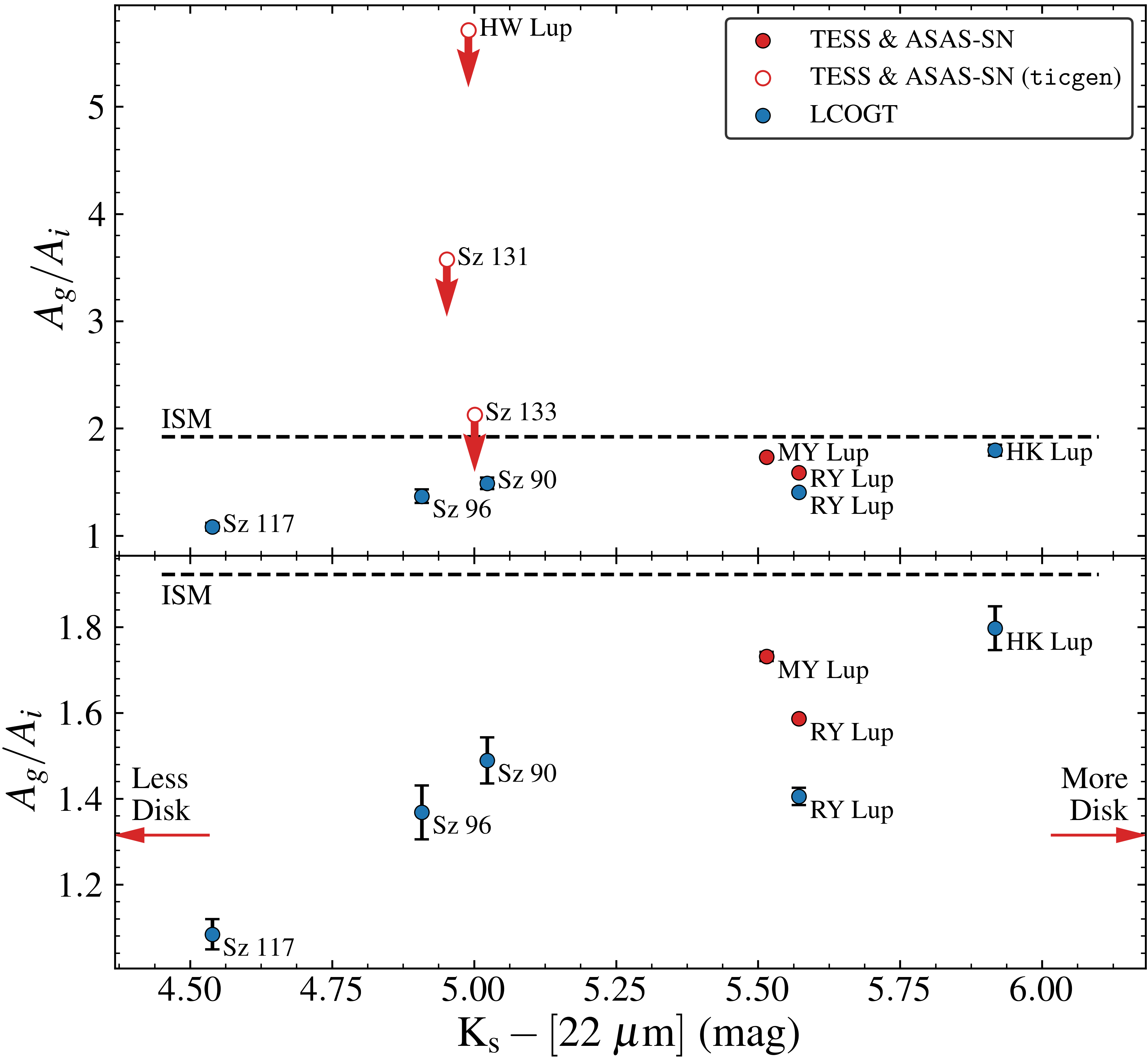}
        \caption{\emph{Top Panel:} The dependence of the $A_g/A_i$ extinction ratio on the source's IR excess. IR Excess is an indicator of excess emission due to a hot inner disk. We see a potential relation between the two, though additional data is needed. Slopes for HW Lup, Sz 131, and Sz 133 are shown as open points to reflect the possible unreliability of the values (see text). \emph{Bottom Panel:} A zoomed view of the dippers with directly measured references fluxes.}
        \label{fig:redden-ir}
    \end{figure*}

    \subsection{Multi-Band Variability and Dust Properties}\label{sec:Dust}

    Simultaneous observations of young stellar variability in two or more band-passes provides a means of differentiating between different phenomena (e.g., flares, accretion, circumstellar dust). In the case of obscuration and reddening by circumstellar dust, it serves as an indicator of the grain size distribution \citep[e.g.,][]{Herbst1994,Bouvier1999,Huang2019}. The relative change (slope) in color (reddening) and magnitude (dimming)  (Equation \ref{eq:color-mag}) depends on the characteristic grain size $d$ and composition. Dust dominated by small grains will produce steeper slopes (more reddening for a given amount of dimming) compared to dust dominated by larger grains, which in the limit of $d \gg \lambda$ will produce no reddening at all. The grain size distribution is affected by grain growth in the disks, and the slope of the trend in a color--magnitude plot is indicative of that history \citep[e.g.,][]{Guo2018}. UX Orionis stars exhibit reddening up to a point where the direct line of sight to the star is almost entirely obscured, causing scattered light that is comparatively blue to dominate, switching the sign of the slope \citep{Bibo1990,Grinin1994}. In contrast, the lack of reddening (and in fact slight bluing) by the repetitive dipper AA Tau has been explained by grains $>1\mu$m \citep{Bouvier1999}. Note that these plots are insensitive to zero-point offsets in magnitudes, since all values are differential.

    Five Lupus dippers (HW Lup, RY Lup, MY Lup, Sz 131, Sz 133) were monitored simultaneously by the \emph{TESS} mission and ASAS-SN. The \emph{TESS} band-pass $T$ spans 600--1000 nm and is centered on the Cousins $I$-band \citep{Ricker2014}. Figure \ref{fig:TESS} shows the $g$-band ASAS-SN and \emph{TESS} data, and the change in the $g-T$ color with $g$. Linear least-$\chi^2$ fits are shown as red lines. Based on the reddening $R$ values in \citet{Yuan2013} and \citet{Stassun2019}, the expected slope of reddening due to interstellar dust is 0.38.

    Light curves and $g-i$ versus $i$ reddening diagrams for the five stars (RY Lup, Sz 90, Sz 96, HK Lup, and Sz 117) observed with LCOGT are shown in Figure \ref{fig:lco}. In this case, the expected slope from interstellar dust is 0.48. In all cases the slope is less than, and sometimes significantly less than, the expected interstellar dust value. Only one star, RY Lup, was observed by all three telescope networks, but not all simultaneously. The \emph{TESS} and LCOGT observations are not contemporaneous, with the former ending about 33 days before the beginning of the latter. The LCOGT slope (0.29) and \emph{TESS}--ASAS-SN slope (0.25) are roughy consistent, with the smaller \emph{TESS}--ASAS-SN slope (larger $A_T$ compared to $A_i$) consistent with the extension of the \emph{TESS} bandpass to shorter wavelengths than the SDSS $i$ band filter. Values of $A_g/A_i$ and $A_g/A_T$, derived from these slops, are included in Table \ref{tab:lup_members}.

    We can find no obvious correlation between the slope of the color--magnitude variation from our \emph{TESS}--ASAS-SN or LCOGT observations and other properties of our disks such as the relative luminosity, the minimum wavelength of the infrared excess \citep{Merin2008}, the disk inclination or physical extent of dust and gas (if resolved) or the estimated dust/gas masses \citep{Ansdell2016b,Ansdell2018b}. However, we do see a possible correlation between the reddening slope and the infrared excess of the disk (WISE $22$ $\mu$m emission) relative to the photosphere (2MASS $\text{K}_\text{s}$ emission). HK Lup, MY Lup, and RY Lup have highest values of $A_g/A_i$ and ostensibly the smallest grains and also the largest infrared excess (Figure \ref{fig:redden-ir}). The system with the lowest $22$ $\mu$m excess, Sz 117, also has little or no reddening, with $A_i/A_g$ near unity. We have included our \emph{TESS} measurements of $A_g/A_i$ for HW Lup, Sz 131, and Sz 133, but we use open symbols and arrows for them in Figure \ref{fig:redden-ir} since the reference flux estimated using {\tt ticgen} may not be reliable. Excluding these objects, the Spearman correlation coefficient of the data indicates a 97\% probability that the two parameters, $A_g/A_i$ and $\text{K}_\text{s}-[22\:\mu\text{m}]$, are correlated.

    A correlation between extinction coefficients and infrared excess could be explained if either (a) coalescence of a fixed mass of dust into larger grains produced both a flatter slope and less emitting area (and hence less emission), or (b) the growth of grains also proceeded contemporaneously with clearing of the disk. The two scenarios are mutually exclusive. With our small sample size (11 systems) and limited spectrophotometric data set we are unable to choose between these two possible explanations; more observations of a larger sample are clearly needed.



\section{Summary}\label{sec:Summary}

We have identified 11 dipper stars in the Lupus molecular cloud using ASAS-SN observations and a series of quantitative selection criteria that identified YSO members of Lupus, variable Lupus YSOs, and Lupus YSOs that varied in a manner consistent with dippers identified in other star-forming regions \citep{Cody2014}. All eleven stars are pre-main sequence, lying above the theoretical zero-age main sequence in a color-magnitude diagram. Their positions in the diagram are consistent with expectations for Lupus-age ($\sim$2 Myr) low-mass ($\lesssim$1 $M_{\odot}$) stars, but we are unable to estimate individual ages or masses due both to the lack of spectra and uncertainties in the effect of spots on theoretical models. All 11 dippers have infrared excesses consistent with a ``full" T Tauri disk, as has been observed for other dipper stars \citep[e.g.,][]{Ansdell2016a}. Our findings expand the scope of the dipping phenomenon to yet another star-forming environment and epoch.

We used multi-wavelength time-series photometry of these dipper stars, combining ASAS-SN, \emph{TESS}, and LCOGT observations, to measure the reddening by the occulting dust and thus infer something about the grain size. In all cases the amount of reddening was less, and thus the grain sizes larger, than that expected from ISM-like dust. Moreover, we find a tentative trend of decreasing reddening, and hence increasing grain size, with decreasing infrared excess, or equivalently more advanced evolutionary stages of the disk. This is consistent with a picture of grain growth with time in protoplanetary disks \citep{Testi2014}. These results need to be confirmed with larger and more in-depth studies, but highlight the synergy of ground- and space-based observations in the effort to better understand the structure and evolution of circumstellar disks and the planetary systems they spawn.



\section*{Acknowledgements}

We thank the Las Cumbres Observatory and its staff for its continuing support of the ASAS-SN project. LCOGT observations were performed as part of DDT award 2019B-003 to EG. ASAS-SN is supported by the Gordon and Betty Moore Foundation through grant GBMF5490 to the Ohio State University and NSF grant AST-1515927. Development of ASAS-SN has been supported by NSF grant AST-0908816, the Mt. Cuba Astronomical Foundation, the Center for Cosmology and AstroParticle Physics at the Ohio State University, the Chinese Academy of Sciences South America Center for Astronomy (CASSACA), the Villum Foundation, and George Skestos.

JWB acknowledges support from Research Experience for Undergraduate program at the Institute for Astronomy, University of Hawaii--Manoa funded through NSF grant 6104374. He would also like to thank the Institute for Astronomy for their kind hospitality during the course of this project. This research was also supported through Hawaii Space Grant Consortium fellowships. JWB would like to thank Jason Hinkle and Emily Heckle for constructive criticism of the manuscript.

This work was supported by NASA grant 80NSSC19K1717. BJS, CSK, and KZS are supported by NSF grant AST-1907570/AST-1908952. BJS is also supported by NSF grants AST-1920392 and AST-1911074.

EG acknowledges support from NSF award AST-187215. EG conducted part of this research during the Exostar19 program at the Kavli Institute for Theoretical Physics at UC Santa Barbara, which was supported in part by the National Science Foundation under Grant No. NSF PHY-1748958. EG was also supported as a visiting professor to the University of G\"{o}ttingen by the German Science Foundation through DFG Research 644 Unit FOR2544 ``Blue Planets around Red Stars".

CSK and KZS are also supported by NSF grants AST-1515927 and AST-181440. Support for JLP is provided in part by FONDECYT through the grant 1191038 and by the Ministry of Economy, Development, and Tourism's Millennium Science Initiative through grant IC120009, awarded to The Millennium Institute of Astrophysics, MAS. SD acknowledges Project 11573003 supported by NSFC. TAT is supported in part by NASA grant 80NSSC20K0531. PJV is supported by the National Science Foundation Graduate Research Fellowship Program Under Grant No. DGE-1343012. We thank Ethan Kruse for uploading videos of the \textit{TESS} FFIs to YouTube, as these are an invaluable aide when examining the images.

This work has made use of data from the European Space Agency (ESA) mission {\it Gaia} (\url{https://www.cosmos.esa.int/gaia}), processed by the {\it Gaia} Data Processing and Analysis Consortium (DPAC, \url{https://www.cosmos.esa.int/web/gaia/dpac/consortium}). Funding for the DPAC has been provided by national institutions, in particular the institutions participating in the {\it Gaia} Multilateral Agreement.

This publication makes use of data products from the Wide-field Infrared Survey Explorer, which is a joint project of the University of California, Los Angeles, and the Jet Propulsion Laboratory/California Institute of Technology, and NEOWISE, which is a project of the Jet Propulsion Laboratory/California Institute of Technology. WISE and NEOWISE are funded by the National Aeronautics and Space Administration.

This publication makes use of data products from the Two Micron All Sky Survey, which is a joint project of the University of Massachusetts and the Infrared Processing and Analysis Center/California Institute of Technology, funded by the National Aeronautics and Space Administration and the National Science Foundation.

This paper includes data collected by the TESS mission, which are publicly available from the Mikulski Archive for Space Telescopes (MAST). Funding for the TESS mission is provided by NASA's Science Mission directorate.

This work is based on observations made by ASAS-SN. We wish to extend our special thanks to those of Hawaiian ancestry on whose sacred mountain of Haleakal\=a, we are privileged to be guests. Without their generous hospitality, the observations presented herein would not have been possible.



\bibliographystyle{mnras}
\input{output.bbl}






\bsp	
\label{lastpage}

\end{document}